# Scaled indium oxide transistors fabricated using atomic layer deposition

Mengwei Si,[1,2] Zehao Lin,[1] Zhizhong Chen,[1] Xing Sun,[3] Haiyan Wang,[3] Peide D. Ye[1]

[1]School of Electrical and Computer Engineering and Birck Nanotechnology Center, Purdue University, West Lafayette, Indiana 47907, United States

[2]Department of Electronic Engineering, Shanghai Jiao Tong University, Shanghai 200240, China

[3]School of Materials Engineering, Purdue University, West Lafayette, IN 47907, United States

*Address correspondence to: yep@purdue.edu (P.D.Y.)

**Abstract**

In order to continue to improve integrated circuit performance and functionality, scaled transistors with short channel lengths and low thickness are needed. But the further scaling of silicon-based devices and the development of alternative semiconductor channel materials that are compatible with current fabrication processes is challenging. Here we report atomic-layer-deposited indium oxide transistors with channel lengths down to 8 nm, channel thicknesses down to 0.5 nm and equivalent dielectric oxide thickness down to 0.84 nm. Due to the scaled device dimensions and low contact resistance, the devices exhibit high on-state currents of 3.1 A/mm at a drain voltage of 0.5 V and a transconductance of 1.5 S/mm at a drain voltage 1 V. Our devices are a promising alternative channel material for scaled transistors with back-end-of-line processing compatibility.

The scaling of complementary metal–oxide–semiconductor (CMOS) technology has been the driving force in the advancement of modern integrated circuits over the past few decades. Enhancing gate electrostatic control to improve immunity to short channel effects (SCEs) has, in particular, been a key strategy in the development of aggressively scaled transistor technology. This includes the development of high-k/metal gate technology for equivalent oxide thickness (EOT) scaling, as well as ultrathin body, fin and stacked nanosheet channel transistors; stacked nanosheet transistors are currently being adopted by the semiconductor industry (following FinFET technology) beyond the 3 nm technology node[1]. To further scale length dimensions, while maintaining good drive current, it is critical to suppress SCEs. This can be achieved using increased numbers of thinner stacked channels. However, the performance of conventional semiconductor based transistors decreases rapidly below a thickness of 3 nm for silicon and 10 nm for InGaAs.

Two-dimensional (2D) semiconductors are an alternate channel material with nanoscale thicknesses and higher mobilities at monolayer or few-layer thickness compared to conventional semiconductors[2–11]. However, 2D materials suffer from a lack of high-quality large-area CMOS-compatible growth techniques. It is also difficult to form dielectrics on their van der Waals surfaces. In addition, the materials are difficult to dope and suffer from high contact resistances induced at the Schottky metal/semiconductor contacts.

Oxide semiconductors — and amorphous indium gallium zinc oxide (IGZO) in particular — are leading semiconducting channel materials in thin-film transistors (TFTs) for flat-panel display applications[12]. But despite being a mature technology for high-volume manufacturing, oxide semiconductors are rarely considered as channel materials for scaled high-performance transistors.

This due to their low charge carrier mobility of about 10 cm$^2$/V·s and the fact that when used in mass production they typically require channel thicknesses of up to several tens of nanometer[13]. However, there has been interest in the use of oxide semiconductor transistors in CMOS back-end-of-line (BEOL) for monolithic 3D integration applications[14–21]. In particular, an atomic layer deposition (ALD) based oxide semiconductor and device technology was developed with channel thicknesses down to sub-1 nm and high field-effective mobility ($\mu_{FE}$) >100 cm$^2$/V·s (ref [20,22–25]). The wafer-scale homogenous and conformal ALD indium oxide (In$_2$O$_3$) thin film has a low thermal budget of 225 °C and an atomically smooth surface. The ability to deposit conformal films on 3D structures by ALD could be useful for 3D integration applications, including 3D BEOL integration and 3D vertical NAND.

ALD deposited In$_2$O$_3$ has a number of advantages for use in transistor applications. The layer-by-layer self-limited growth mechanism of ALD[26–29] enables an atomically smooth surface, so that nanometre-thick uniform thin films can be achieved that suppress roughness related carrier scattering and deterioration of band structure. In addition, the charge neutrality level (CNL) of In$_2$O$_3$ lies deep inside the conduction band so a high electron density exists in atomically thin channels providing a more forgiving metal to semiconductor low-resistive contact due to Fermi level pinning[24], thus leading to the high on-state currents in transistors that use nanometre thick channels.

In this Article, we report high-performance In$_2$O$_3$ transistors fabricated using ALD. Our devices have channel lengths ($L_{ch}$) down to 8 nm, channel thicknesses ($T_{ch}$) as low as 0.5 nm and equivalent dielectric oxide thicknesses (EOT) down to 0.84 nm. The scaling and low contact resistance of the devices enable them to achieve on-state currents ($I_{D,max}$) of 3.1 A/mm at drain voltages ($V_{DS}$) of 0.5 V and transconductance values ($g_m$) of 1.5 S/mm at $V_{DS}$ of 1 V. Our devices offer promising performance compared to devices based on conventional semiconducting materials (such as Si and GaAs), 2D semiconductors, and other oxide semiconductors, particularly at the ultrathin scale of 1–3.5 nm.

**Results and Discussion**

Fig. 1a illustrates the schematic diagram of the ALD $In_2O_3$ transistor in this work. The gate stack includes 40 nm Ni as gate metal and $HfO_2$ as gate dielectric unless otherwise specified, 0.5 to 3.5 nm $In_2O_3$ as semiconducting channels. $In_2O_3$ by ALD exhibits an atomically smooth surface with surface roughness as low as 0.16 nm,[22] measured by atomic force microscopy (AFM), being beneficial to the thickness scaling of ALD $In_2O_3$. The device has a channel width ($W_{ch}$) of 2 μm for $L_{ch}$ equal or above 40 nm while a $W_{ch}$ of 0.6 μm for $L_{ch}$ below 40 nm. The $W_{ch}$ was accurately defined by e-beam lithography and dry etching as shown in Fig. S1b. A reduced $W_{ch}$ is used to suppress the self-heating effect. Ni by e-beam evaporation is used as source/drain electrodes. The device fabrication process is discussed in great details in methods section. Fig. 1b shows the high-angle annular dark field scanning transmission electron microscopy (HAADF-STEM) with energy-dispersive x-ray spectroscopy (EDX) mapping of a representative $In_2O_3$ transistor with $L_{ch}$ of 8 nm, $T_{ch}$ of 3.5 nm and hafnium oxide thickness ($T_{ox}$) of 3 nm, highlighting Ni/In/O elements. EDX mapping with Ni/In/Hf/O elements is shown in Fig. S1a. $L_{ch}$ defined as the distance between source/drain Ni electrodes is measured to be 8 nm in this device. Fig. 1c presents the HAADF-STEM with EDX mapping images on a $W/HfO_2/Al_2O_3/In_2O_3/Ni$ stack, with $In_2O_3$ thicknesses from 0.7 nm to 1.5 nm. The thicknesses of $In_2O_3$ are determined by both AFM measurements and TEM images here similar to that in previous work.[24] Fig. 1d and 1e show the $I_D$-$V_{GS}$ and $I_D$-$V_{DS}$ characteristics of a representative ALD In$_2$O$_3$ transistor with L$_{ch}$ of 1 μm, T$_{ch}$ of 1.2 nm and 10 nm HfO$_2$/1 nm Al$_2$O$_3$ as gate insulator. The device exhibits high on/off ratio > 10$^{10}$ due to the relatively wide bandgap of In$_2$O$_3$ and negligible hysteresis due to the high-quality oxide/semiconductor interface. This device has a subthreshold (SS) of 130.4 mV/dec at V$_{DS}$ of 1 V, which can be furtherly reduced by EOT scaling and proper interface engineering.[22,23]

To further improve the performance of ALD In$_2$O$_3$ transistors, device scaling is performed including channel length scaling, EOT scaling and channel thickness scaling, where EOT scaling and channel thickness scaling are essential to enhance the gate electrostatic control to improve the immunity to short channel effects. T$_{ch}$ scaling down to 0.5 nm is achieved in this work as shown in the I$_D$-V$_{GS}$ characteristics with T$_{ch}$ of 0.8 nm, 0.65 nm and 0.5 nm in Figs. 2a-2c. The corresponding I$_D$-V$_{DS}$ characteristics are shown in Figs. 2d-2f. The devices have SS of 109.0 mV/dec, 114.3 mV/dec and 108.2 mV/dec at V$_{DS}$ of 1 V for T$_{ch}$ of 0.8 nm, 0.65 nm and 0.5 nm, respectively. Here, 5 nm HfO$_2$ is used as gate dielectric. As we can see, even with T$_{ch}$ of 0.5 nm, well-behaved transfer characteristics with on/off ratio over 7 orders are achieved. A functional transistor made of a 3D semiconducting channel material and with T$_{ch}$ as low as 0.5 nm has never been reported before because of two reasons. First, it is very challenging to achieve atomically smooth and uniform semiconducting film by conventional thin-film deposition techniques such as sputtering, chemical vapor deposition (CVD), etc., resulting in the degradation of electrical performance due to carrier scattering and the change of band structures due to roughness. Second, it is difficult to form a good metal/semiconductor contact with low contact resistivity on ultrathin semiconducting film. Fermi level pinning problem owing to the metal-induced gap states (MIGS) results in the Schottky barrier at metal/semiconductor interface, which is more severe in ultrathin film due to the quantum confinement effects. ALD In$_2$O$_3$ is found to overcome these two challenges at nanometer scale. First, the layer-by-layer self-limiting growth mechanism ensures the atomically smooth surface and uniform film. Second, the CNL of bulk $In_2O_3$ aligns about 0.4 eV above $E_C$, so that thick $In_2O_3$ behaves like a conducting oxide.[30] As a result, Fermi level is pinned above $E_C$ for metal/$In_2O_3$ contact, leading to a low contact resistance below 0.1 $\Omega \cdot mm$ even at nanometer scale. The contact resistance was extracted by transmission line method (TLM) as previously reported in Ref. 20. Note that, the above property is fundamental when searching for a 3D semiconductor with high performance at ultrathin channel. It is expected that a 3D semiconductor with CNL aligns far above $E_C$ (or far below valence band edge as $E_V$ for p-type devices) is necessary to achieve high-performance transistor with ultrathin channel at nanometer scale. Figs. 2g-2i present the maximum drain current ($I_{D,max}$), transconductance ($g_m$) and field-effective mobility ($\mu_{FE}$) scaling metrics. $\mu_{FE}$ is calculated using the maximum $g_m$ at low drain bias. The on-state performance degrades rapid below 1 nm, as shown in Fig. 2. Devices with $T_{ch}$ of 0.4 nm has no detectable drain currents as shown in Fig. S2a. And $I_{D,max}$ reduces exponentially when $T_{ch}$ decreases linearly from 0.8 nm to 0.5 nm, as shown in Fig. S2b due to the impact of quantum confinement on the band structure of $In_2O_3$ thin film. The drain current scaling metrics is found to deviate from 1/L as shown in Fig. 2g, indicating that it is more difficult to accumulate carriers in long channels, most likely due to the percolation mechanism.[31] Additional data on SS and $V_T$ scaling metrics are shown in Fig. S3.

A second approach to enhance the gate electrostatic control is EOT scaling. Here, we demonstrate high-performance ALD $In_2O_3$ transistors with EOT scaled down to 0.84 nm. Fig. 3a shows the C-V characteristics of the gate stack capacitor (Ni/3 nm $HfO_2$/3.5 nm $In_2O_3$/Ni), fabricated together with $In_2O_3$ transistors with same gate stack. EOT is calculated to be 0.84 nm, using $C_{ox}=\epsilon_0\epsilon_{SiO2}/EOT$, where $\epsilon_{SiO2}$ is 3.9 as dielectric constant of $SiO_2$, $\epsilon_0$ is $8.85\times10^{-14}$ F/cm as vacuum permittivity and $C_{ox}$ is measured from C-V measurement at high $V_{GS}$ at low frequency. These numbers confirm a high-quality bulk gate oxide and oxide/semiconductor interface. The doping concentration ($N_D$) of $In_2O_3$ can be extracted from $1/C^2$ versus voltage characteristics according to $N_D = \frac{2}{q\epsilon_S\epsilon_0 d(1/C^2)/dV}$,[32] where q is the elementary charge, $\epsilon_S$ is the relative dielectric constant of $In_2O_3$, which is about 8.9.[33] Fig. S4 shows the $1/C^2$ versus voltage characteristics obtained from Fig. 3a. From the slope of $1/C^2$ versus voltage characteristics, $N_D$ can be estimated to be $9.0 \times 10^{19}$ /cm$^3$. Fig. 3b shows the $I_D$-$V_{DS}$ characteristics of an $In_2O_3$ transistor with $L_{ch}$ of 1 μm, $T_{ch}$ of 3.5 nm and EOT of 0.84 nm, showing well-behaved drain current saturation. Figs. 3c and 3d show the $I_D$-$V_{GS}$ and $I_D$-$V_{DS}$ characteristics of an $In_2O_3$ transistor with $L_{ch}$ of 50 nm, $T_{ch}$ of 3.5 nm and EOT of 0.84 nm. The device has SS of 387.3 mV/dec at $V_{DS}$ of 0.7 V. The relatively large SS and apparent drain-induced-barrier-lowering (DIBL) is partly due to the gate leakage current in highly scaled EOT. Maximum $I_D$ of 2.4 A/mm is achieved at a low $V_{DS}$ of 0.7 V. Drain current saturation is not achieved due to $V_{GS}$-$V_T$ is much larger than $V_{DS}$, so that the pinch-off condition is not fulfilled. Figs. 3e and 3f summarize the $I_{ON}$ and $g_m$ scaling metrics of $In_2O_3$ transistors with $L_{ch}$ from 1 μm down to 50 nm, $T_{ch}$ of 3.5 nm and $T_{ox}$ from 3 nm to 5 nm. Each data point represents the average of at least 5 devices. The small error bar in these plots demonstrates that the ALD based $In_2O_3$ transistors are highly uniform. $I_{ON}$ is extracted at $V_{GS}$-$V_T$=1 V and $V_{DS}$=1 V. Maximum $g_m$ is extracted at $V_{DS}$=1 V. $I_{ON}$ and $g_m$ are found to be improved significantly by EOT scaling. A high average $I_{ON}$ of 1.2 A/mm is achieved at $L_{ch}$ of 50 nm and at $V_{GS}$-$V_T$=1 V and $V_{DS}$=1 V. A high average $g_m$ of 1.5 S/mm is achieved at $L_{ch}$ of 50 nm and at $V_{DS}$=1 V. There values are the best among known reported oxide semiconductor transistors.

Figs. 4a and 4b show the $I_D$-$V_{GS}$ and $I_D$-$V_{DS}$ characteristics of ALD $In_2O_3$ transistors with $L_{ch}$ of 8 nm, $T_{ch}$ of 2.5 nm, and EOT of 0.84 nm. A record high $I_{D,max}$ of 3.1 A/mm is achieved due to the scaled device dimension and low contact resistance, where $I_D > 3$ A/mm is achieved on oxide semiconductor transistors. The $I_{D,max}$ and $g_m$ scaling metrics with $L_{ch}$ from 1 μm down to 8 nm of $In_2O_3$ transistors are shown in Figs. 4c and 4d. A lower $V_{DS}$ of 0.5 V is used for shorter channel length to avoid the self-heating effect due to high power density in ultra-scaled device area. The corresponding $V_{DS}$ at different $L_{ch}$ is marked in the figure. Considering $V_{DS}$ of 0.5 V and $I_D$ of 3.1 A/mm, $R_C$ can be estimated to be less than 0.08 $\Omega \cdot$ mm.

The performance of scaled ALD $In_2O_3$ transistors in this work are benchmarked with state-of-the-art high-performance transistors with ultrathin channel, such as 2D transistors and oxide semiconductor transistors, using figure of merits of $I_{ON}$, $g_m$, $R_C$ and mobility versus channel thickness (the data is summarized in supplementary Table I). ALD $In_2O_3$ transistors exhibit the largest $I_{D,max}$ in the range of 1-3.5 nm and largest $g_m$ below 3.5 nm, among all known semiconductor thin film to the authors' best knowledge, as shown in Figs. 5a and 5b. Such high-performance material and device properties are mainly contributed by the low contact resistance benefiting from the unique CNL alignments in $In_2O_3$, as shown in Fig. 5c, and the high mobility in nanometer scale between 1-3.5 nm, also benefiting from the atomically thickness control of ALD, as shown in Fig. 5d.

**Conclusion**

In summary, we have reported an ALD based oxide semiconductor transistor technology that shows promising on-state currents — in comparison to both established and emerging material technologies — when channel thicknesses are fabricated in the range of approximately 1 to 3.5 nm. Our approach takes advantage of the self-limiting growth mechanism of ALD and the unique band structure of $In_2O_3$. The conformal deposition on 3D structures by ALD also the potential to create new opportunities for 3D integration, such as BEOL compatible transistors for monolithic 3D integration and semiconducting channels for 3D vertical NAND.

## Methods

**Device Fabrication.** The device fabrication process is similar to previous work.[22] The device fabrication process started with solvent cleaning of p+ Si substrate with thermally grown 90 nm $SiO_2$. A bi-layer photoresist lithography process (PMGI SF9 + AZ1518) was then applied for the sharp lift-off of 40 nm Ni gate metal by e-beam evaporation. This step is critical to avoid sidewall metal coverage, so that high-quality ALD gate dielectric can be formed for EOT scaling. $HfO_2$ with various thicknesses as gate insulator were deposited by ALD at 200 °C with $[(CH_3)_2N]_4Hf$ (TDMAHf) and $H_2O$ as Hf and O precursors. $In_2O_3$ thin films with various thicknesses were deposited by ALD at 225 °C using $(CH_3)_3In$ (TMIn) and $H_2O$ as In and O precursors. $N_2$ is used as carrier gas at a flow rate of 40 sccm. The base pressure at $N_2$ flow rate of 0 sccm is 169 mTorr while base pressure at $N_2$ flow rate of 40 sccm is 437 mTorr. Concentrated hydrochloric acid was employed for the channel isolation. S/D ohmic contacts were formed by e-beam evaporation of Ni in two steps to avoid the difficulty on sub-10 nm lift-off process. It is difficult to form sub-10 nm channel length by one step e-beam lithography because of the proximity effect that causing the electron back scattering to the channel region. Therefore, a two-step e-beam lithography process was adopted by the formation of source electrode first and then the drain electrode. A wide range of distances between source and drain electrodes are defined in the masks so that both short and long channel length can be achieved. Then, a second step ICP dry etching using $BCl_3$/Ar plasma was used to accurately define the channel width. The devices were annealed in $O_2$ at 250 °C for 4 min to further improve the performance.

**Material Characterization.** The thickness of the $In_2O_3$ was determined together by AFM, TEM and ellipsometry. AFM measurement was done with a Veeco Dimension 3100 atomic force microscope system. TEM lamella samples were prepared with Helios G4 UX Dual Beam SEM.

FEI TALOS F200X operated at 200 kV equipped with super-X electron-dispersive X-ray spectroscopy was used for HAADF-STEM imaging.

**Device Characterization.** Electrical characterization was carried out with a Keysight B1500 system and with a Cascade Summit probe station in dark and $N_2$ environments at room temperature and at atmosphere.

**Data availability.** The data that support the plots within this paper and other findings of this study are available from the corresponding author upon reasonable request.

**Acknowledgements**

This work was supported in part by the Semiconductor Research Corporation (SRC) nCore Innovative Materials and Processes for Accelerated Compute Technologies (IMPACT) Center and in part by the Air Force Office of Scientific Research (AFOSR) and SRC/Defense Advanced Research Projects Agency (DARPA) Joint University Microelectronics Program (JUMP) Applications and Systems-driven Center for Energy Efficient integrated Nano Technologies (ASCENT) Center.

**Author Contributions**

P.D.Y. and M.S. conceived the idea and proposed the ALD $In_2O_3$ scaling research. M.S. developed the atomic layer deposition process of $In_2O_3$ as high-performance oxide semiconductor. M.S. and Z.L. did the device fabrication, electrical measurement and analysis on thickness and EOT scaling of ALD $In_2O_3$ devices. Z.L. and M.S. conducted the channel length scaling of ALD $In_2O_3$ devices down to 8 nm. Z.C., X.S. and H.W. performed the STEM and EDX measurements. M.S. and P.D.Y. co-wrote the manuscript and all authors commented on it.

**Financial Interest Statement**

The authors declare no competing financial interest.

**Supplementary Information**

Additional details for HAADF-STEM cross-sectional image with EDX element mapping (O, Ni, In and Hf), I-V characterstics of ALD $In_2O_3$ transistors with $T_{ch}$ of 0.4 nm, SS and $V_T$ scaling metrics, C-V characterization of the gate stack capacitor, detailed information on the benchmarking are in the supplementary information.

**Figure Captions**

**Figure 1 | Schematic diagram, TEM images and I-V characteristics of ALD $In_2O_3$ transistors.** **a,** Schematic diagram of an ALD $In_2O_3$ transistor. **b,** HAADF-STEM cross sectional image with EDX element mapping of an $In_2O_3$ transistor with $L_{ch}$ of 8 nm, $T_{ch}$ of 3.5 nm and 3 nm $HfO_2$ as gate insulator, capturing the 8 nm channel length. **c,** HAADF-STEM cross-sectional images with EDX element mapping and AFM measurements of $In_2O_3$ transistors with W/$HfO_2$/$Al_2O_3$/$In_2O_3$/Ni gate stack with $T_{ch}$ of from 0.7 nm to 1.5 nm. **d,** $I_D$-$V_{GS}$ and **e,** $I_D$-$V_{DS}$ characteristics of a representative ALD $In_2O_3$ transistor with $L_{ch}$ of 1 μm, $T_{ch}$ of 1.2 nm and 10 nm $HfO_2$/1 nm $Al_2O_3$ as gate insulator. The device exhibits high on/off ratio $> 10^{10}$ and negligible hysteresis due to the relatively wide bandgap semiconductor and high-quality oxide/semiconductor interface.

**Figure 2 | Thickness scaling of ALD In$_2$O$_3$ down to 0.5 nm.** I$_D$-V$_{GS}$ characteristics of ALD In$_2$O$_3$ transistors with L$_{ch}$ of 40 nm, T$_{ch}$ of **a,** 0.8 nm, **b,** 0.65 nm and **c,** 0.5 nm and 5 nm HfO$_2$ as gate insulator. I$_D$-V$_{DS}$ characteristics of the same ALD In$_2$O$_3$ transistors as in Figs. 2a-2c with L$_{ch}$ of 40 nm, T$_{ch}$ of **d,** 0.8 nm, **e,** 0.65 nm and **f,** 0.5 nm and 5 nm HfO$_2$ as gate insulator. **g,** I$_{D,max}$, **h,** g$_m$ and **i,** μ$_{FE}$ scaling metrics of ALD In$_2$O$_3$ transistors with different T$_{ch}$ and with 5 nm HfO$_2$ as gate insulator at V$_{DS}$ of 1 V. Each data point represents the average of at least 5 devices. Well-behaved transfer and output characteristics with on/off ratio > 10$^7$ are achieved with channel thickness down to 0.5 nm. The impact of Schottky barrier at metal/semiconductor interface on output characteristics is clearly observed with T$_{ch}$ is below 1 nm due to the effect of quantum confinement on the band structure of ultrathin In$_2$O$_3$ film.

**Figure 3 | EOT scaling of ALD In$_2$O$_3$ transistors down to sub-1 nm. a,** C-V measurement of the gate stack capacitor (Ni/3 nm HfO$_2$/3.5 nm In$_2$O$_3$/Ni) at 2 kHz, fabricated together with the ALD In$_2$O$_3$ transistor on the same chip. EOT of 0.84 nm is achieved, suggesting a high quality HfO$_2$/In$_2$O$_3$ interface by ALD. **b,** I$_D$-V$_{DS}$ characteristics of an In$_2$O$_3$ transistor with L$_{ch}$ of 1 μm, T$_{ch}$ of 3.5 nm and EOT of 0.84 nm, showing well-behaved drain saturation due to the high drain bias. A high drain current of 835 μA/μm is achieved at this long L$_{ch}$ of 1 μm. **c,** I$_D$-V$_{GS}$ and **d,** I$_D$-V$_{DS}$ characteristics of an In$_2$O$_3$ transistor with L$_{ch}$ of 50 nm, T$_{ch}$ of 3.5 nm and EOT of 0.84 nm, at V$_{DS}$ of 0.05 V and 0.7 V. The relatively low on/off ratio is because of the gate leakage current at off-state, which may be further improve by threshold voltage tuning. **e,** I$_{ON}$ and **f,** g$_m$ scaling metrics of In$_2$O$_3$ transistors with L$_{ch}$ from 1 μm to 50 nm, T$_{ch}$ of 3.5 nm and T$_{ox}$ of 3, 3.5 and 5 nm. I$_{ON}$ is extracted at V$_{GS}$-V$_T$=1V and V$_{DS}$=1 V. g$_m$ is extracted at V$_{DS}$=1 V. Each data point represents

the average of at least 5 devices. High $I_{ON}$ of 1.2 A/mm at $V_{GS}-V_T=1$ V and $V_{DS}=1$ V and $g_m$ of 1.5 S/mm at $V_{DS}=1$ V are achieved.

**Figure 4 | Channel length scaling of ALD In$_2$O$_3$ transistors down to 8 nm. a,** $I_D$-$V_{GS}$ and **b,** $I_D$-$V_{DS}$ characteristics of In$_2$O$_3$ transistors with $L_{ch}$ of 8 nm, $T_{ch}$ of 2.5 nm, and EOT of 0.84 nm. **c,** $I_{D,max}$ and **d,** $g_m$ scaling metrics of best-performance In$_2$O$_3$ transistors with $L_{ch}$ from 1 μm to 8 nm with $T_{ch}$ of 2.5 nm. Lower voltages at shorter channel devices are used to avoid the impact of self-heating on devices.

**Figure 5 | Benchmarking of ALD In$_2$O$_3$ with other ultrathin semiconductors.** Comparison of **a,** $I_{D,max}$, **b,** $g_m$, **c,** $R_C$ and **d,** mobility versus channel thickness characteristics with other high-performance oxide semiconductor (doped-In$_2$O$_3$) devices by sputtering and 2D semiconductor devices (MoS$_2$, WS$_2$, BP). The data used in this figure is listed in supplementary Table I. ALD In$_2$O$_3$ demonstrates best performance in terms of $I_{D,max}$, $g_m$, $R_C$ and mobility in 1-3.5 nm range compared to all known semiconducting materials to the authors' best knowledge.

**References**

1. Loubet, N. *et al.* Stacked nanosheet gate-all-around transistor to enable scaling beyond FinFET. in *IEEE Symposium on VLSI Technology* T230–T231 (IEEE, 2017). doi:10.23919/VLSIT.2017.7998183.

2. Lingming Yang *et al.* High-performance $MoS_2$ field-effect transistors enabled by chloride doping: Record low contact resistance (0.5 kΩ·µm) and record high drain current (460 µA/µm). in *IEEE Symposium on VLSI Technology* 192–193 (IEEE, 2014). doi:10.1109/VLSIT.2014.6894432.

3. Krasnozhon, D., Dutta, S., Nyffeler, C., Leblebici, Y. & Kis, A. High-frequency, scaled $MoS_2$ transistors. in *IEEE Int. Electron Devices Meet.* 703–706 (IEEE, 2015). doi:10.1109/IEDM.2015.7409781.

4. Liu, Y. *et al.* Pushing the Performance Limit of Sub-100 nm Molybdenum Disulfide Transistors. *Nano Lett.* **16**, 6337–6342 (2016). doi:10.1021/acs.nanolett.6b02713.

5. Yang, L. M. *et al.* Few-layer black phosporous PMOSFETs with $BN/Al_2O_3$ bilayer gate dielectric: Achieving $I_{on}$= 850 µA/µm, $g_m$= 340 µS/µm, and $R_c$=0.58 kΩ·µm. in *IEEE Intl. Electron Devices Meet.* 127-130 (IEEE, 2016). doi:10.1109/IEDM.2016.7838354.

6. Si, M., Yang, L., Du, Y. & Ye, P. D. Black phosphorus field-effect transistor with record drain current exceeding 1 A/mm. in *2017 75th Annual Device Research Conference (DRC)* 1–2 (IEEE, 2017). doi:10.1109/DRC.2017.7999395.

7. Li, X. *et al.* High-speed black phosphorus field-effect transistors approaching ballistic limit. *Sci. Adv.* **5**, eaau3194 (2019). doi:10.1126/sciadv.aau3194.

8. Chou, A.-S. *et al.* High On-Current 2D nFET of 390 µA/µm at $V_{DS}$ = 1V using Monolayer CVD $MoS_2$ without Intentional Doping. in *IEEE Symposium on VLSI Technology* TN1.7

(IEEE, 2020). doi:10.1109/VLSITechnology18217.2020.9265040.

9. Shen, P. C. *et al.* Ultralow contact resistance between semimetal and monolayer semiconductors. *Nature* **593**, 211–217 (2021). doi:10.1038/s41586-021-03472-9.

10. Lin, D. *et al.* Scaling synthetic $WS_2$ dual-gate MOS devices towards sub-nm CET. in *IEEE Symposium on VLSI Technology* T3-1 (IEEE, 2021).

11. McClellan, C. J., Yalon, E., Smithe, K. K. H., Suryavanshi, S. V. & Pop, E. High Current Density in Monolayer $MoS_2$ Doped by $AlO_x$. *ACS Nano* **15**, 1587–1596 (2021). doi: 10.1021/acsnano.0c09078.

12. Nomura, K. *et al.* Amorphous Oxide Semiconductors for High-Performance Flexible Thin-Film Transistors. *Jpn. J. Appl. Phys.* **45**, 4303–4308 (2006). doi: 10.1143/JJAP.45.4303.

13. Kamiya, T., Nomura, K. & Hosono, H. Present status of amorphous In–Ga–Zn–O thin-film transistors. *Sci. Technol. Adv. Mater.* **11**, 044305 (2010). doi: 10.1088/1468-6996/11/4/044305.

14. Matsubayashi, D. *et al.* 20-nm-Node trench-gate-self-aligned crystalline In-Ga-Zn-Oxide FET with high frequency and low off-state current. in *IEEE Int. Electron Devices Meet.* 141–144 (IEEE, 2015). doi:10.1109/IEDM.2015.7409641.

15. Li, S. *et al.* Nanometre-thin indium tin oxide for advanced high-performance electronics. *Nat. Mater.* **18**, 1091–1097 (2019). doi:10.1038/s41563-019-0455-8.

16. Li, S., Gu, C., Li, X., Huang, R. & Wu, Y. 10-nm Channel Length Indium-Tin-Oxide transistors with $I_{on}$ = 1860 μA/μm, $G_m$ = 1050 μS/μm at $V_{ds}$ = 1 V with BEOL Compatibility. in *IEEE Int. Electron Devices Meet.* 905–908 (IEEE, 2020). doi:10.1109/IEDM13553.2020.9371966.

17. Samanta, S. *et al.* Amorphous IGZO TFTs featuring Extremely-Scaled Channel Thickness

and 38 nm Channel Length : Achieving Record High $G_{m,max}$ of 125 µS/µm at $V_{DS}$ of 1 V and $I_{ON}$ of 350 µA/µm. in *IEEE Symposium on VLSI Technology* TH2.3 (IEEE, 2020). doi:10.1109/VLSITechnology18217.2020.9265052.

18. Chakraborty, W. *et al.* BEOL Compatible Dual-Gate Ultra Thin-Body W-Doped Indium-Oxide Transistor with $I_{on}$ = 370µA/µm, SS = 73mV/dec and $I_{on}/I_{off}$ ratio > $4x10^9$. in *Symposium on VLSI Technology* TH2.1 (2020). doi: 10.1109/VLSITechnology18217.2020.9265064.

19. Si, M. *et al.* Indium–Tin-Oxide Transistors with One Nanometer Thick Channel and Ferroelectric Gating. *ACS Nano* **14**, 11542–11547 (2020). doi:10.1021/acsnano.0c03978.

20. Si, M., Lin, Z., Charnas, A. & Ye, P. D. Scaled Atomic-Layer-Deposited Indium Oxide Nanometer Transistors With Maximum Drain Current Exceeding 2 A/mm at Drain Voltage of 0.7 V. *IEEE Electron Device Lett.* **42**, 184–187 (2021). doi:10.1109/LED.2020.3043430.

21. Han, K. *et al.* First Demonstration of Oxide Semiconductor Nanowire Transistors: a Novel Digital Etch Technique, IGZO Channel, Nanowire Width Down to ~20 nm, and $I_{on}$ Exceeding 1300 µA/µm. in *IEEE Symposium on VLSI Technology* (IEEE, 2021).

22. Si, M., Charnas, A., Lin, Z. & Ye, P. D. Enhancement-Mode Atomic-Layer-Deposited $In_2O_3$ Transistors With Maximum Drain Current of 2.2 A/mm at Drain Voltage of 0.7 V by Low-Temperature Annealing and Stability in Hydrogen Environment. *IEEE Trans. Electron Devices* **68**, 1075–1080 (2021). doi:10.1109/TED.2021.3053229.

23. Charnas, A., Si, M., Lin, Z. & Ye, P. D. Enhancement-mode atomic-layer thin $In_2O_3$ transistors with maximum current exceeding 2 A/mm at drain voltage of 0.7 V enabled by oxygen plasma treatment. *Appl. Phys. Lett.* **118**, 052107 (2021). doi:10.1063/5.0039783.

24. Si, M. *et al.* Why $In_2O_3$ Can Make 0.7 nm Atomic Layer Thin Transistors. *Nano Lett.* **21**,

500–506 (2021). doi:10.1021/acs.nanolett.0c03967.

25. Si, M., Lin, Z., Chen, Z. & Ye, P. D. First Demonstration of Atomic-Layer-Deposited BEOL-Compatible In$_2$O$_3$ 3D Fin Transistors and Integrated Circuits: High Mobility of 113 cm$^2$/V·s, Maximum Drain Current of 2.5 mA/μm and Maximum Voltage Gain of 38 V/V in In$_2$O$_3$ Inverter. in *IEEE Symposium on VLSI Technology* T2-4 (IEEE, 2021).

26. Mourey, D. A., Zhao, D. A., Sun, J. & Jackson, T. N. Fast PEALD ZnO thin-film transistor circuits. *IEEE Trans. Electron Devices* **57**, 530–534 (2010). doi: 10.1109/TED.2009.2037178.

27. Kim, H. Y. *et al.* Low-Temperature Growth of Indium Oxide Thin Film by Plasma-Enhanced Atomic Layer Deposition Using Liquid Dimethyl(N-ethoxy-2,2-dimethylpropanamido)indium for High-Mobility Thin Film Transistor Application. *ACS Appl. Mater. Interfaces* **8**, 26924–26931 (2016). doi:10.1021/acsami.6b07332.

28. Ma, Q. *et al.* Atomic-Layer-Deposition of Indium Oxide Nano-films for Thin-Film Transistors. *Nanoscale Res. Lett.* **13**, 4 (2018). doi:10.1186/s11671-017-2414-0.

29. Lee, J. *et al.* High mobility ultra-thin crystalline indium oxide thin film transistor using atomic layer deposition. *Appl. Phys. Lett.* **113**, 112102 (2018). doi:10.1063/1.5041029.

30. Robertson, J. & Clark, S. J. Limits to doping in oxides. *Phys. Rev. B* **83**, 075205 (2011). doi:10.1103/PhysRevB.83.075205.

31. Kamiya, T., Nomura, K. & Hosono, H. Electronic Structures Above Mobility Edges in Crystalline and Amorphous In-Ga-Zn-O: Percolation Conduction Examined by Analytical Model. *J. Disp. Technol.* **5**, 462–467 (2009). doi: 10.1109/JDT.2009.2022064.

32. Schroder, D. K. *Semiconductor Material and Device Characterization*. (John Wiley & Sons, Inc., 2006). doi:10.1002/0471749095.

33. Hamberg, I. & Granqvist, C. G. Evaporated Sn-doped In$_2$O$_3$ films: Basic optical properties and applications to energy-efficient windows. *J. Appl. Phys.* **60**, R123–R160 (1986). doi: 10.1063/1.337534.

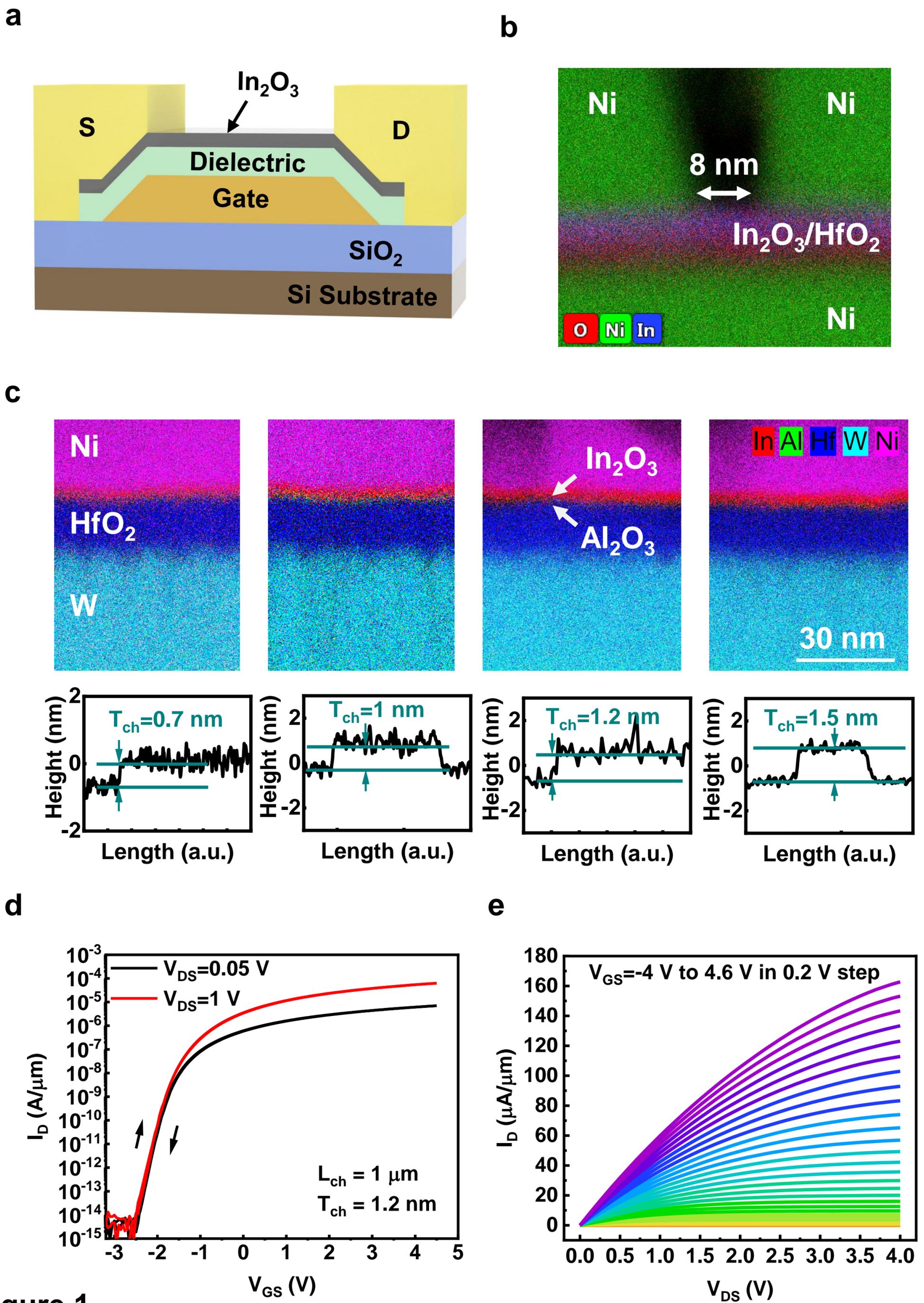

Figure 1

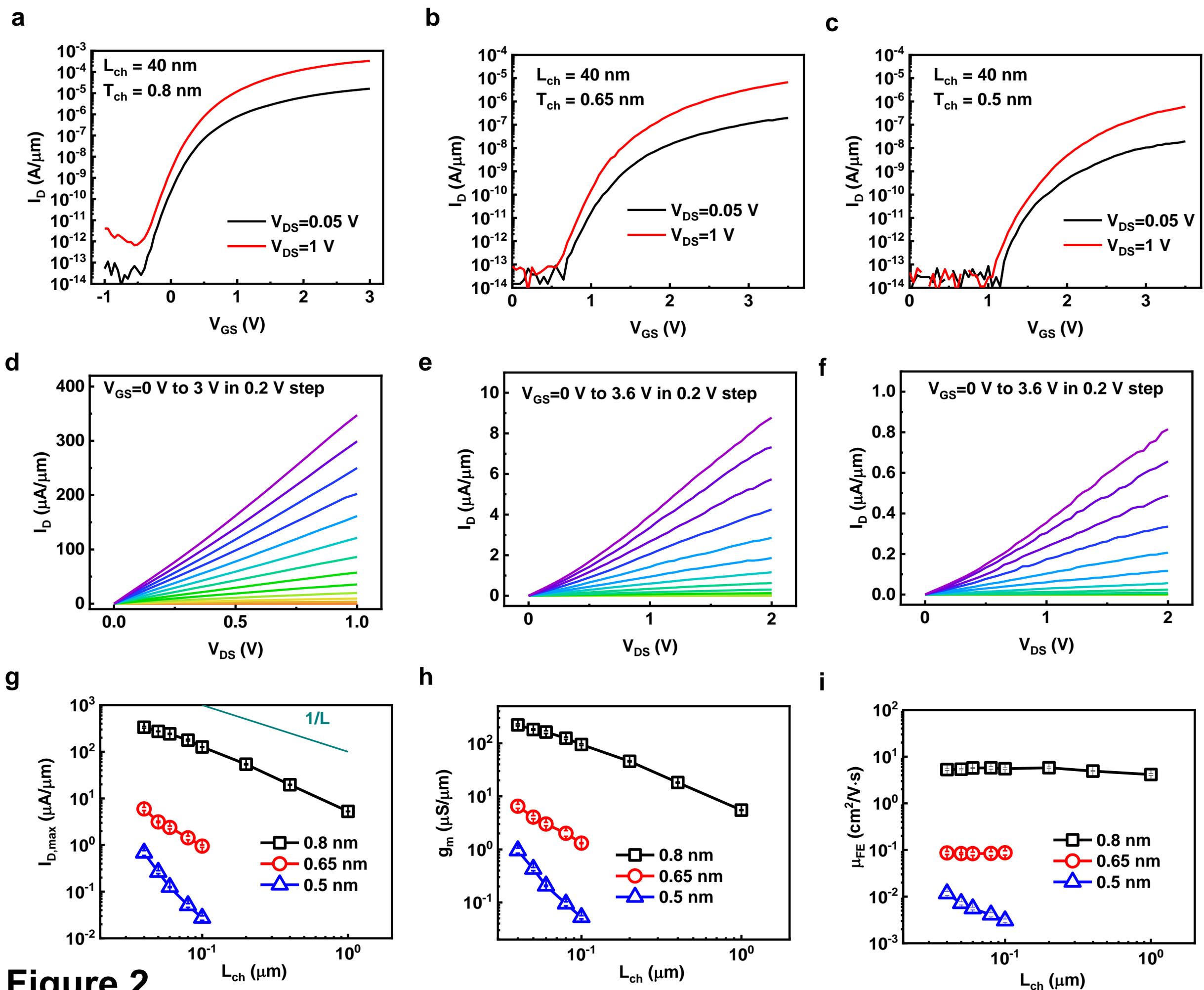

Figure 2

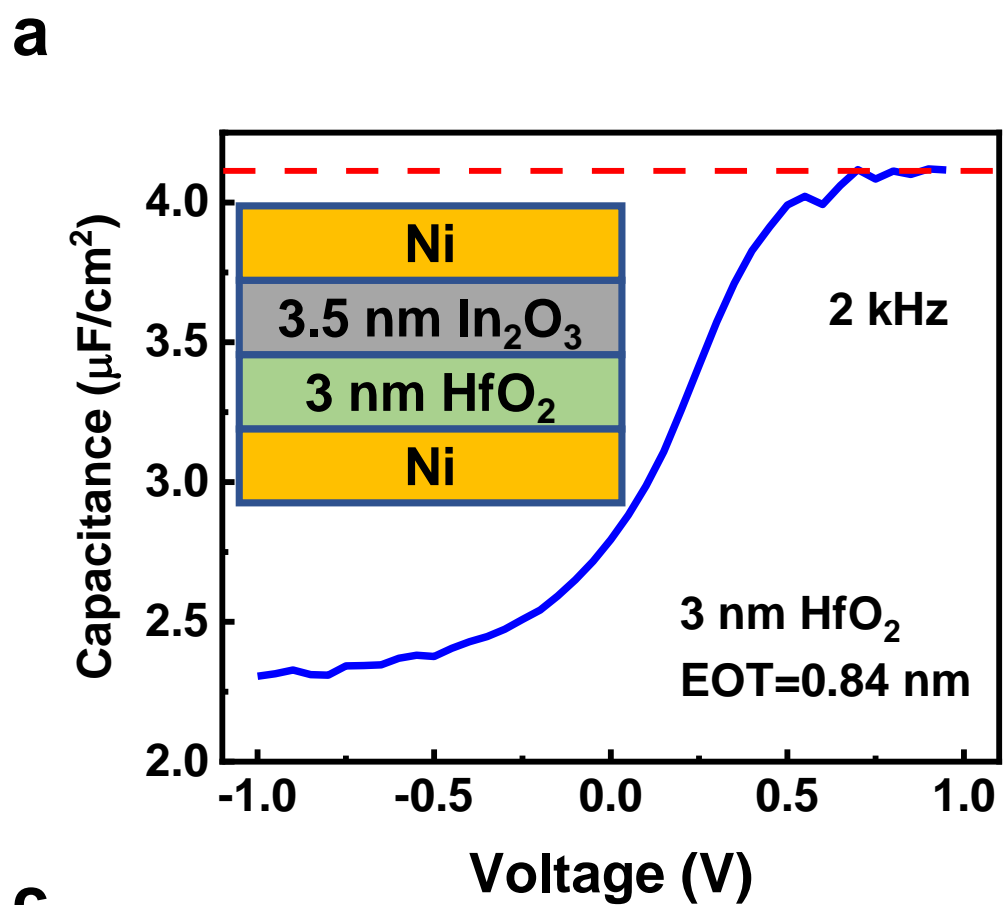
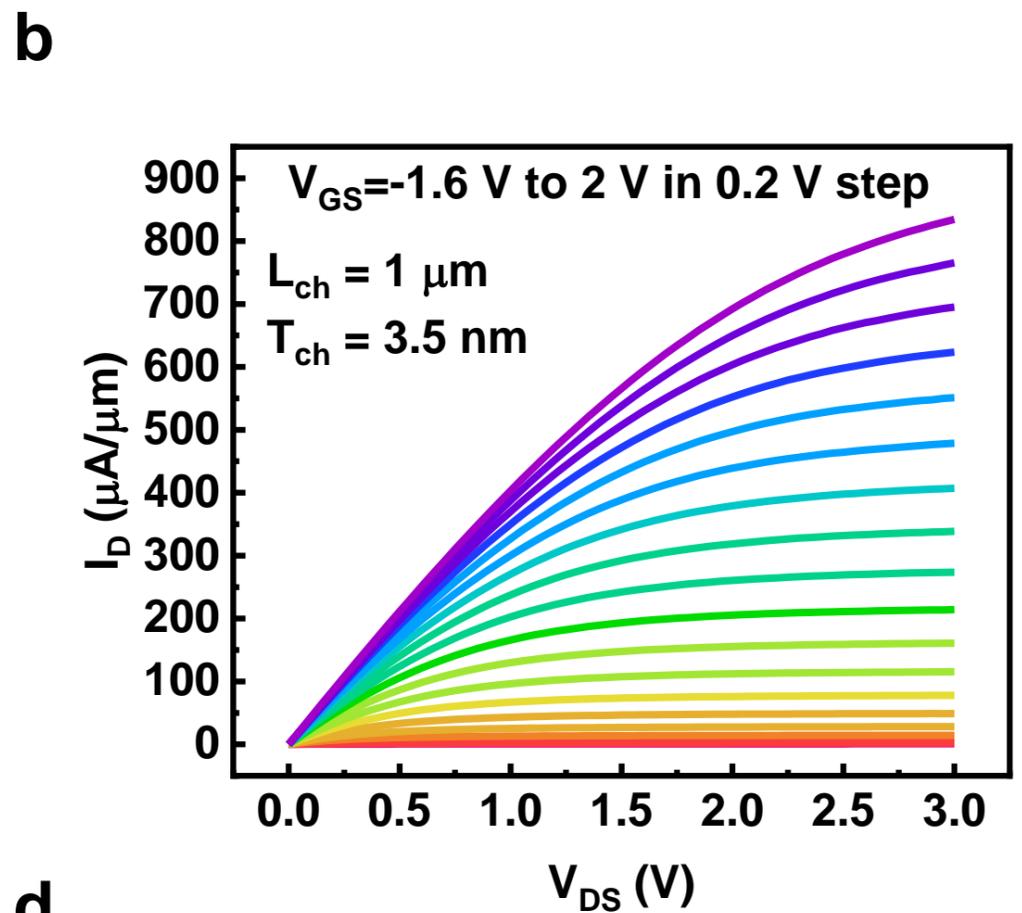
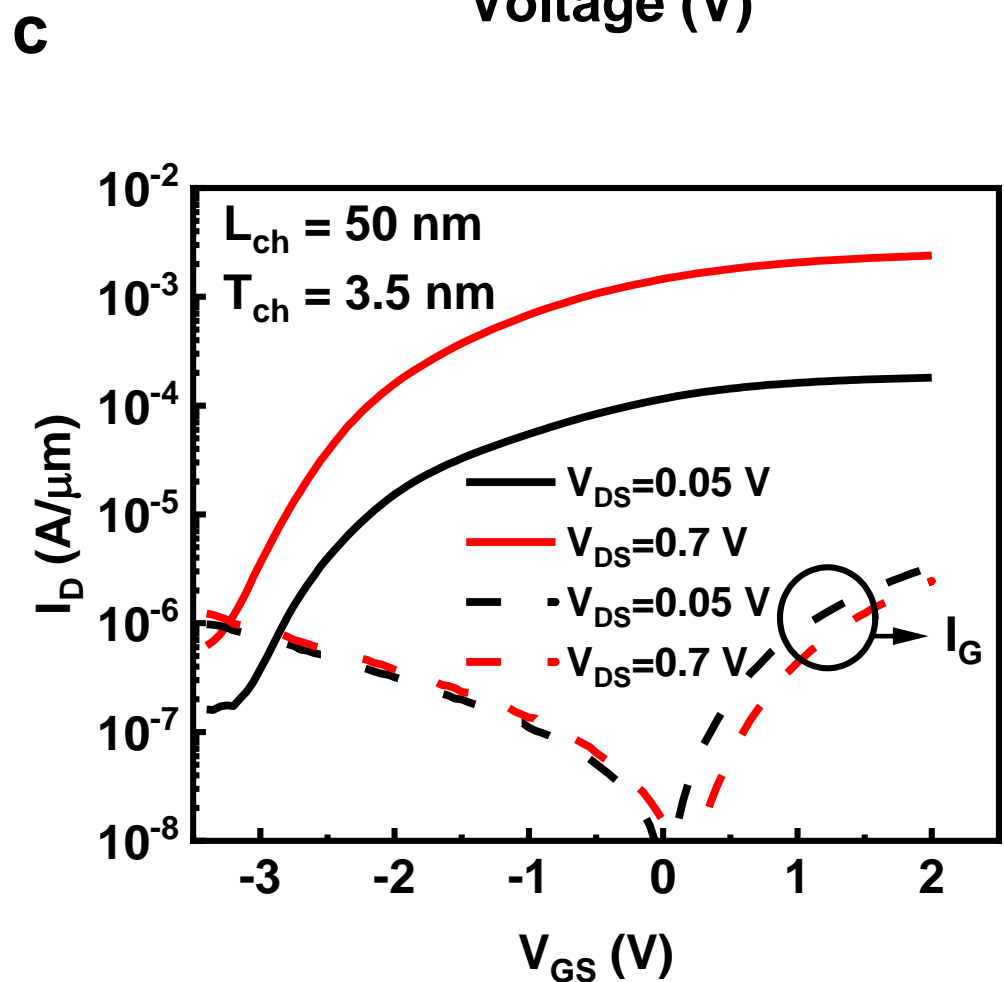
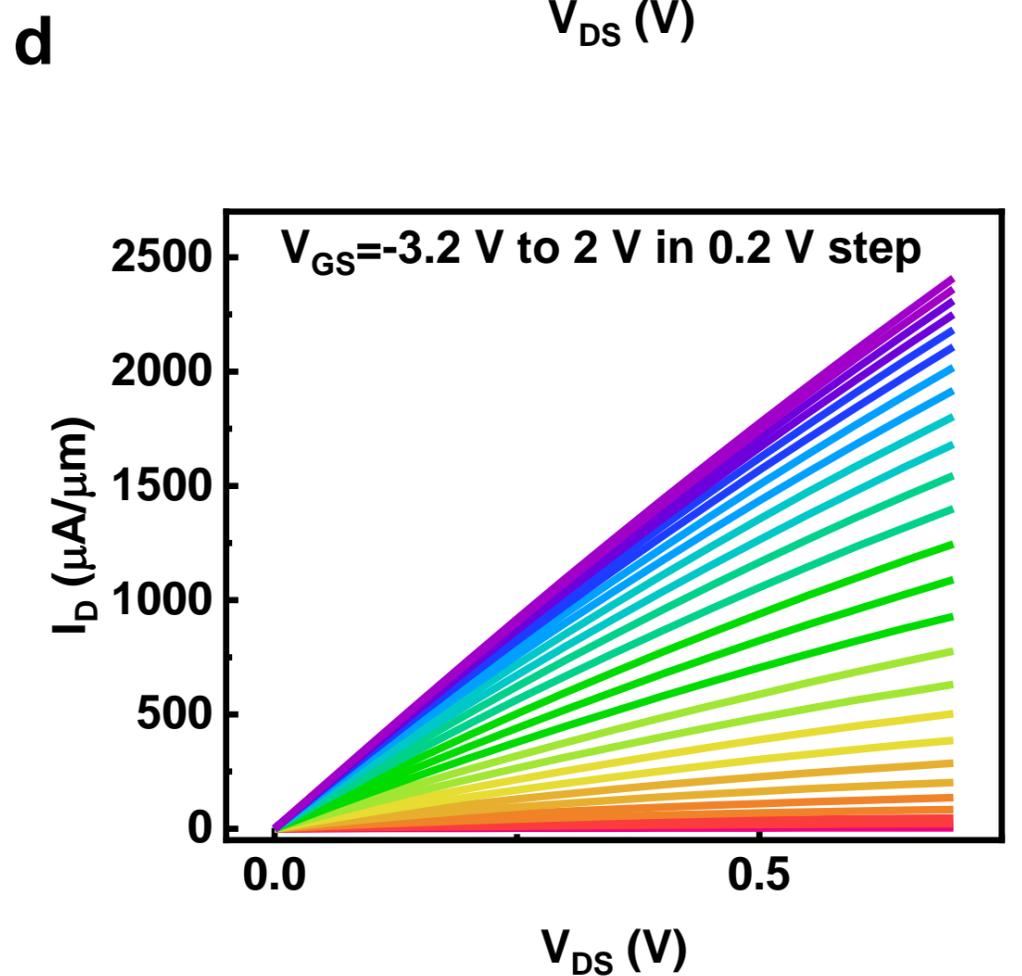
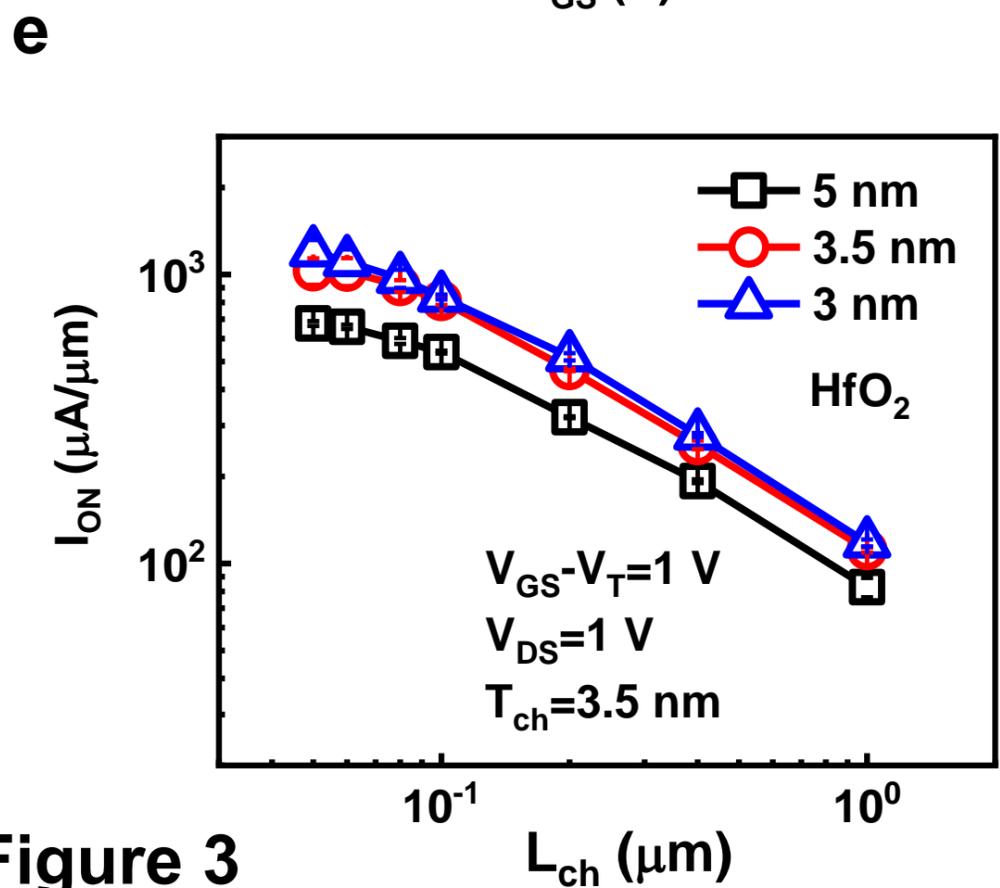
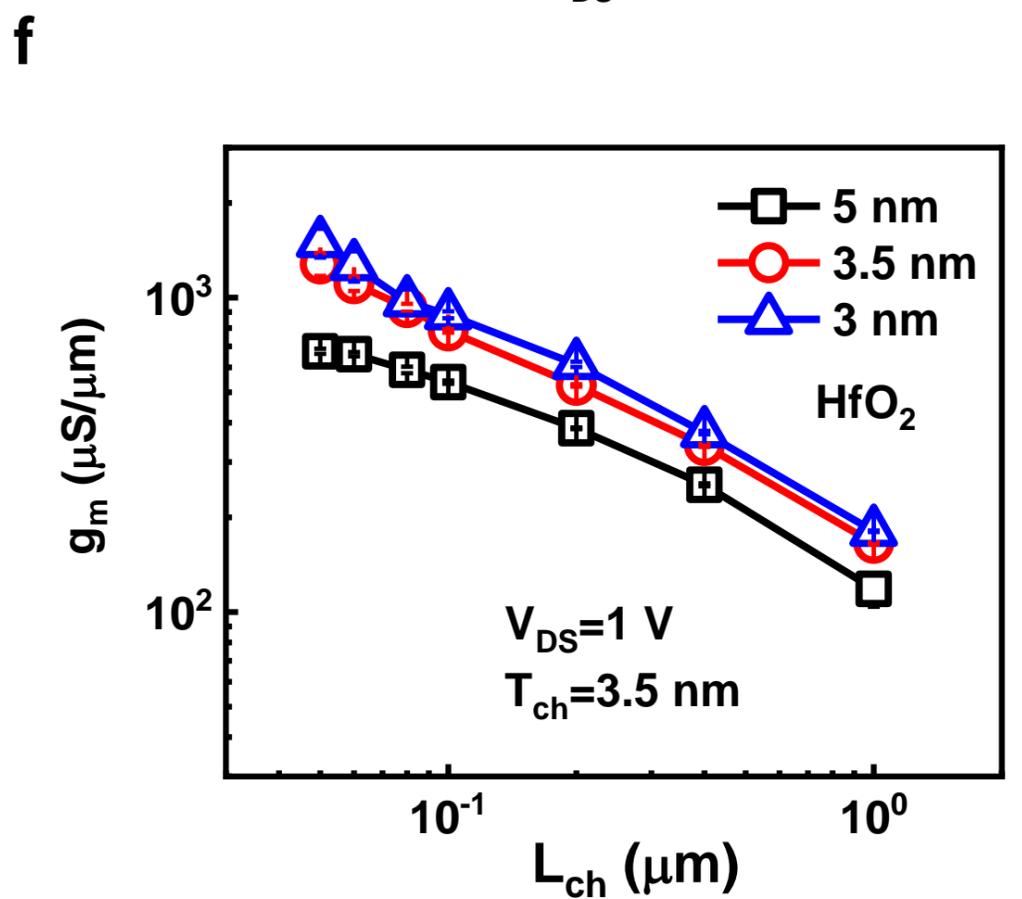

Figure 3

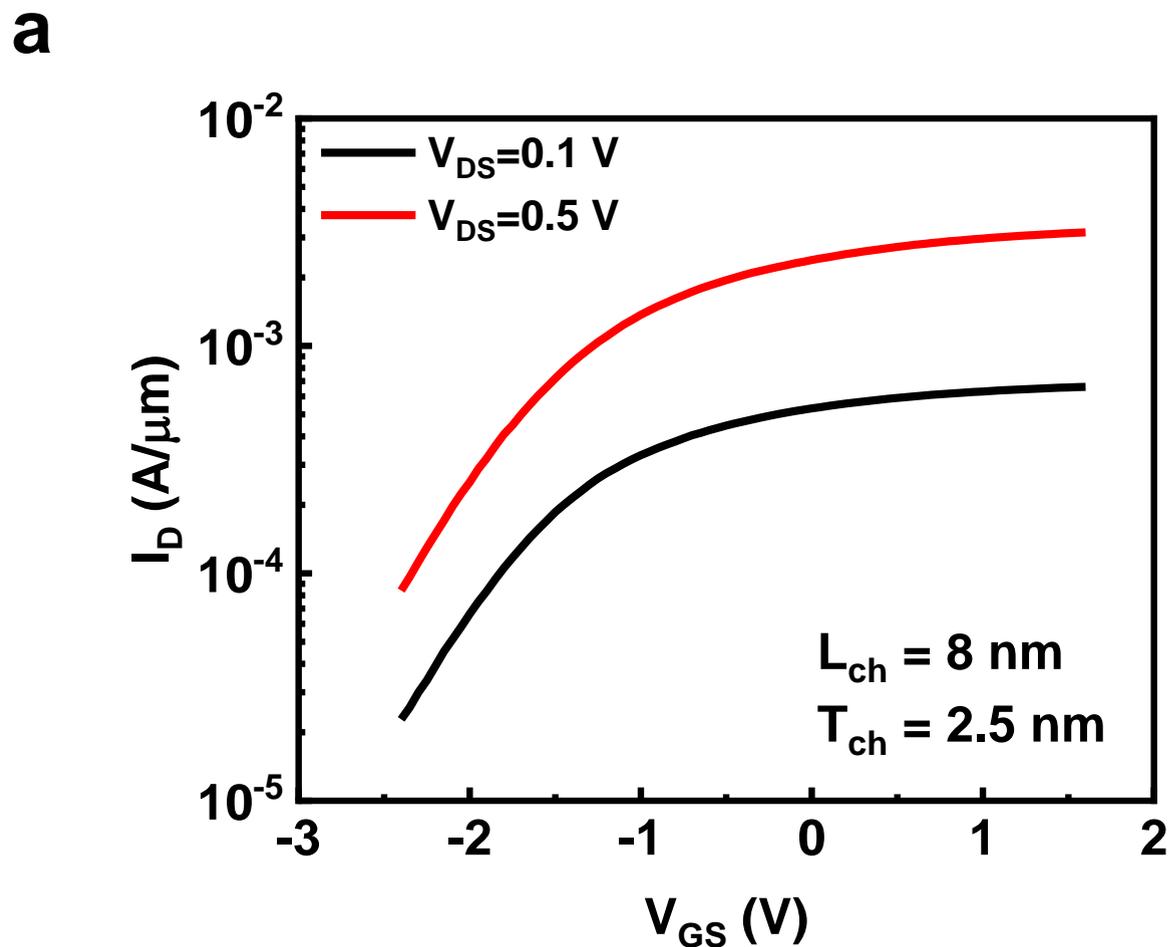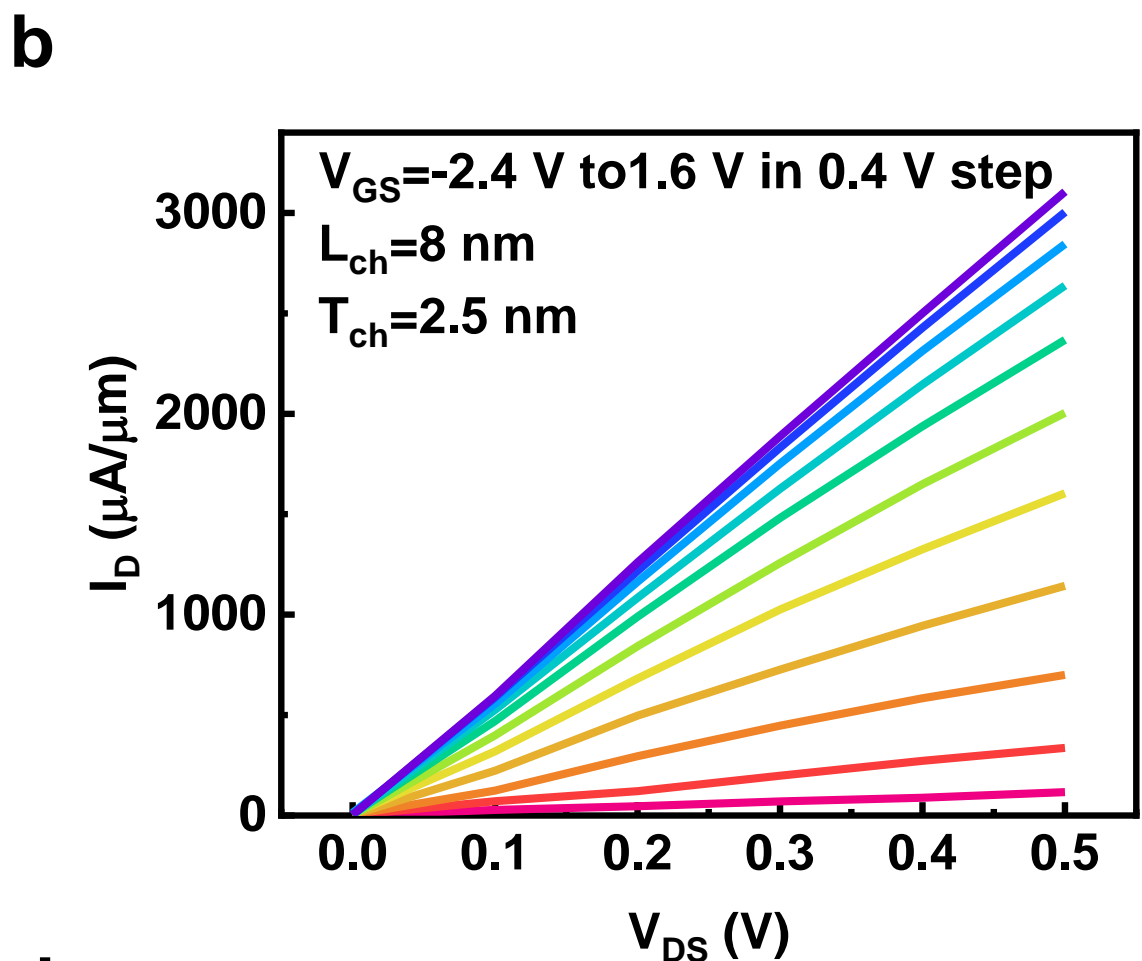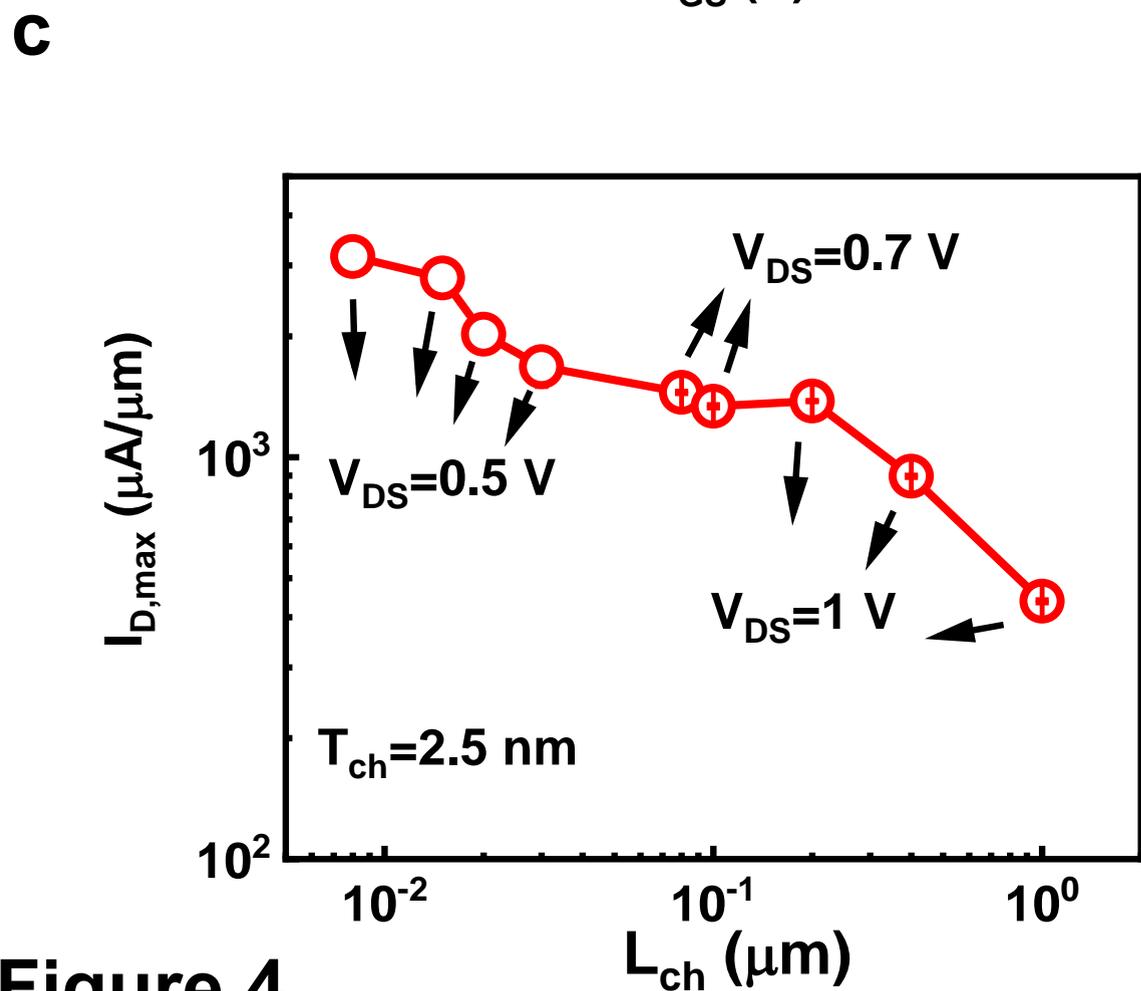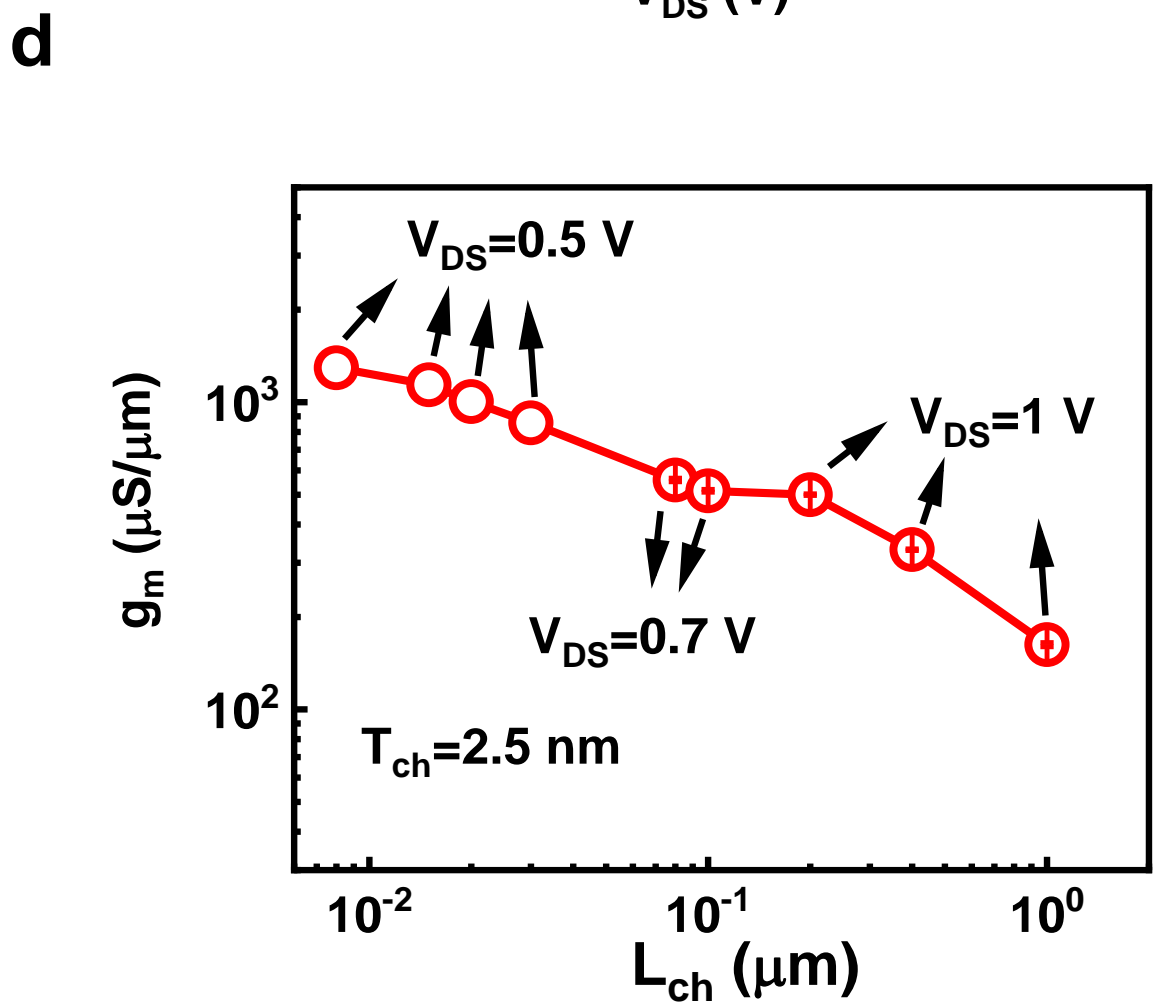

Figure 4

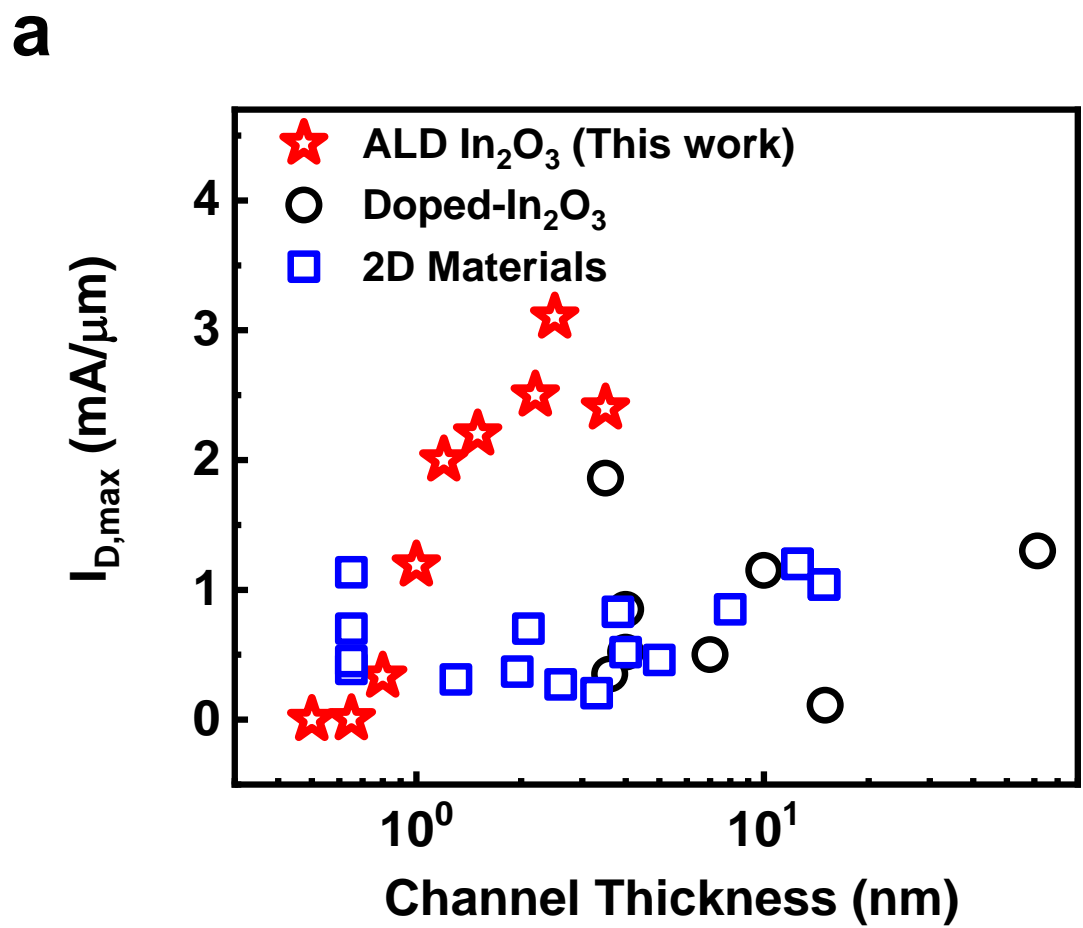
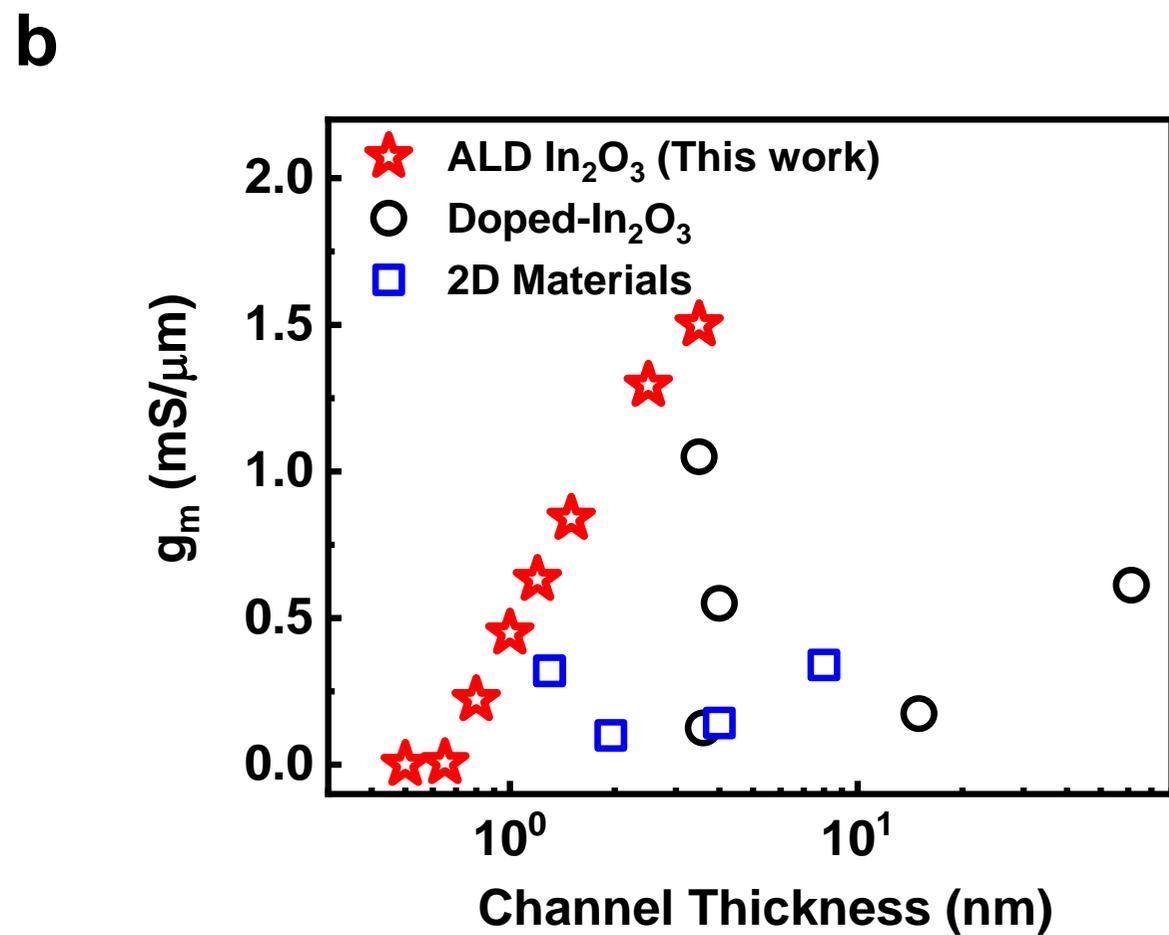
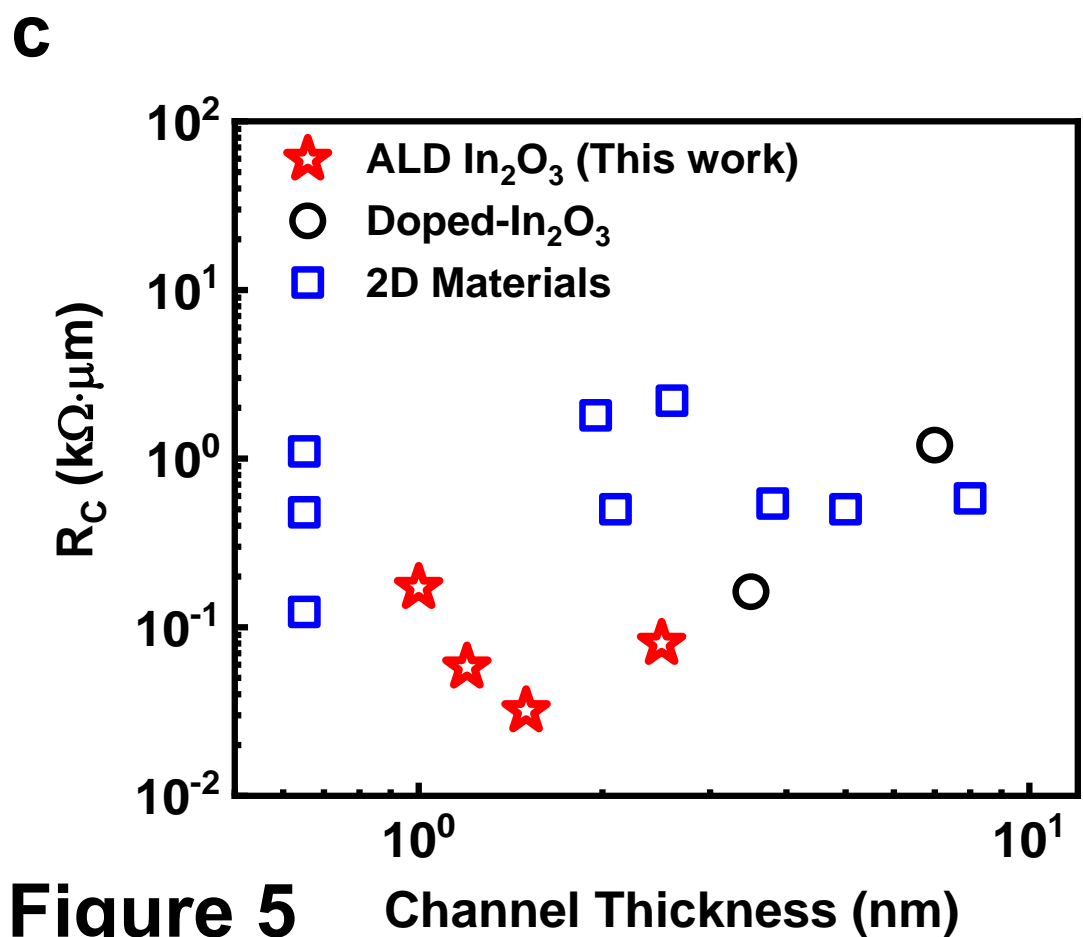
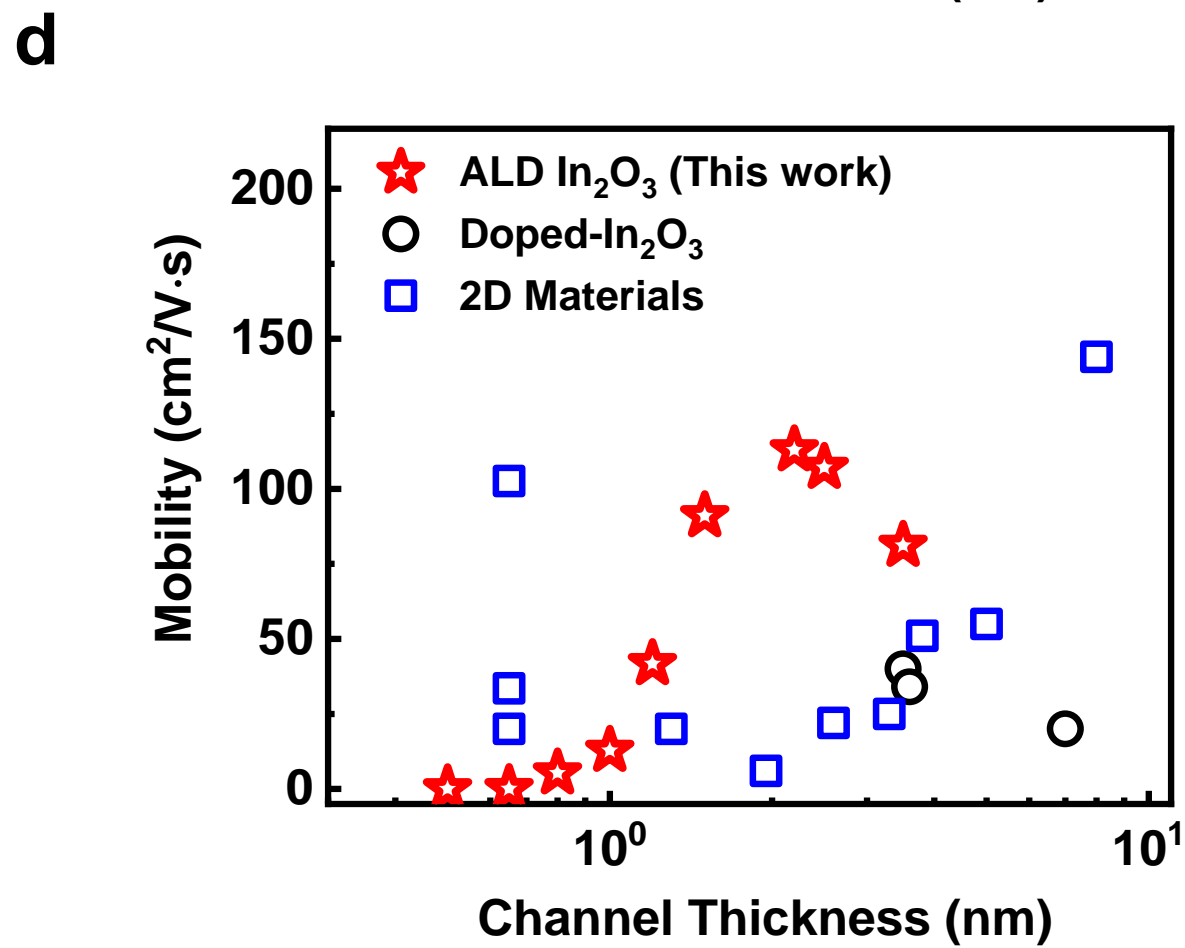

Figure 5

Supplementary Information for:

# Scaled indium oxide transistors fabricated using atomic layer deposition

Mengwei Si,[1,2] Zehao Lin,[1] Zhizhong Chen,[1] Xing Sun,[3] Haiyan Wang,[3] Peide D. Ye[1]

[1]School of Electrical and Computer Engineering and Birck Nanotechnology Center, Purdue University, West Lafayette, Indiana 47907, United States

[2]Department of Electronic Engineering, Shanghai Jiao Tong University, Shanghai 200240, China

[3]School of Materials Engineering, Purdue University, West Lafayette, IN 47907, United States

*Address correspondence to: yep@purdue.edu (P.D.Y.)

# 1. Characteristics of Ultrathin In$_2$O$_3$ Transistors

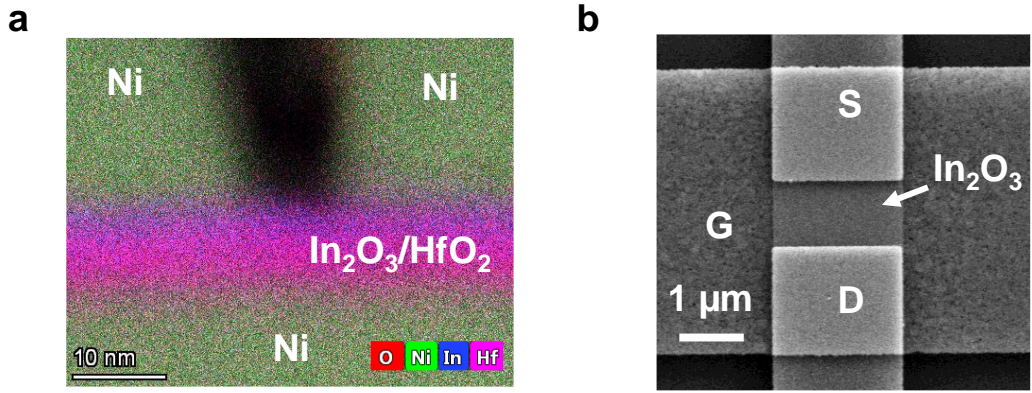

**Figure S1. a,** TEM cross-sectional image with EDX element mapping (O, Ni, In and Hf) of an In$_2$O$_3$ transistor with L$_{ch}$ of 8 nm, T$_{ch}$ of 3.5 nm and 3 nm HfO$_2$ as gate insulator. **b,** SEM image from top view of a typical long channel device with a channel width of 2 μm defined by dry etching.

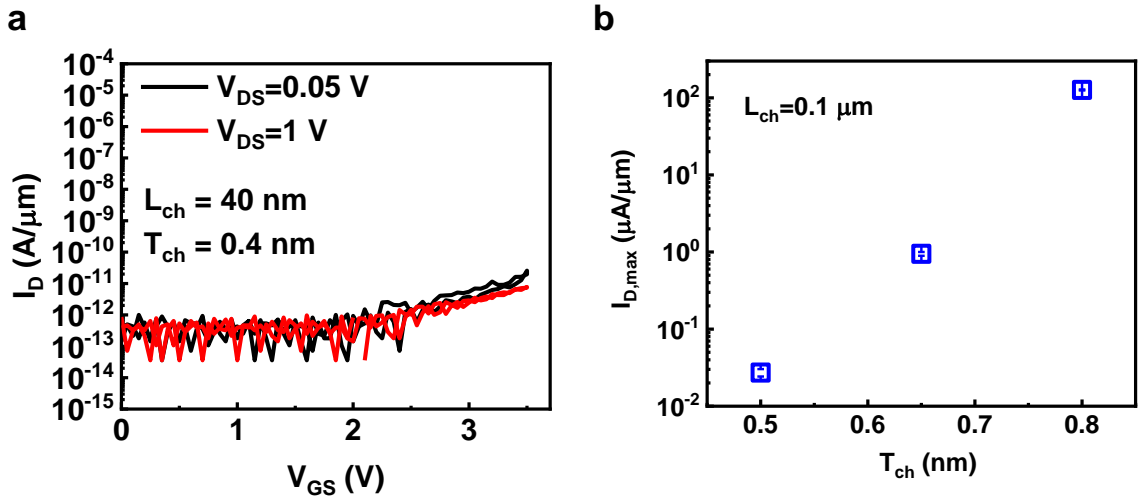

**Figure S2. a,** I$_D$-V$_{GS}$ characteristics of a representative ALD In$_2$O$_3$ transistor with L$_{ch}$ of 40 nm, T$_{ch}$ of 0.4 nm and 5 nm HfO$_2$ as gate insulator. **b,** I$_{D,max}$ versus T$_{ch}$ characteristics for ALD In$_2$O$_3$ transistors extracted from Fig. 2g.

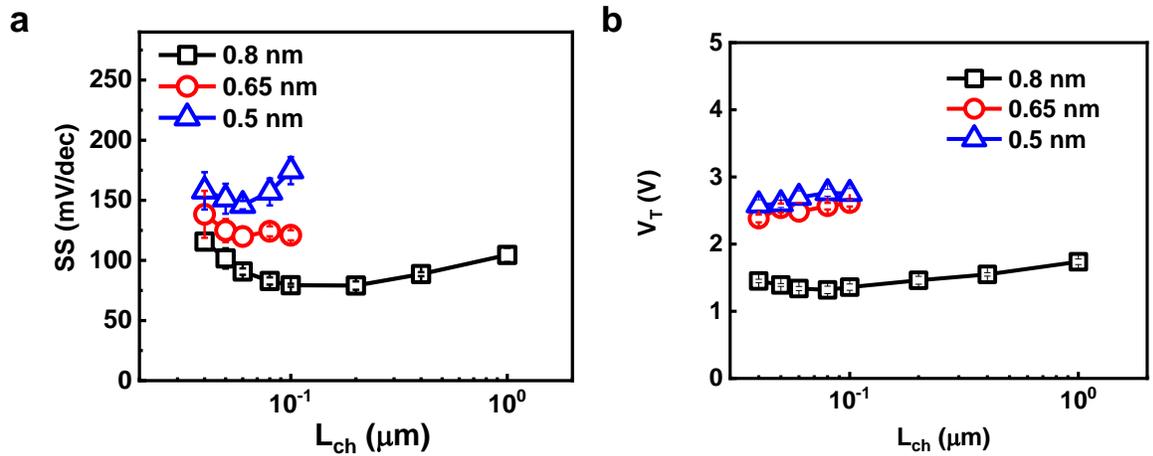

**Figure S3. a,** SS and **b,** $V_T$ scaling metrics of ALD $In_2O_3$ transistors with different $T_{ch}$ and with 5 nm $HfO_2$ as gate insulator.

## 2. C-V Characterization of the Gate Stack Capacitor

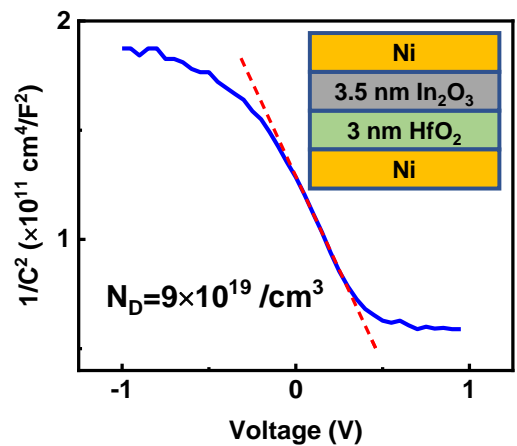

**Figure S4.** $1/C^2$ versus voltage characteristics of a MOS capacitor with Ni/3 nm $HfO_2$/3.5 nm $In_2O_3$/Ni stack.

## 3. Benchmarking of ALD $In_2O_3$ Transistors

Table I. Performance of State-of-the-Art Transistors with Ultrathin Semiconducting Channel

| Material | Thickness (nm) | $L_{ch}$ (nm) | $I_{D,max}$ (A/μm) | $g_m$ (S/μm) | mobility (cm²/V·s) | $R_C$ (kΩ·μm) | Reference |
|---|---|---|---|---|---|---|---|
| ITO | 4 | 40 | 5.20E-04 | 5.50E-04 | - | - | [1] |
| ITO | 4 | 100 | 8.50E-04 | - | - | - | [1] |
| ITO | 10 | 200 | 1.15E-03 | - | - | - | [1] |
| ITO | 3.5 | 10 | 1.86E-03 | 1.05E-03 | 40 | 0.162 | [2] |
| IGZO | 3.6 | 38 | 3.50E-04 | 1.25E-04 | 34 | - | [3] |
| IGZO | 61.3 | 100 | 1.30E-03 | 6.12E-04 | - | - | [4] |
| IGZO | 15 | 27 | 1.10E-04 | 1.74E-04 | - | - | [5] |
| IWO | 7 | 100 | 5.00E-04 | - | 20 | 1.2 | [6] |
| $MoS_2$ | 0.65 | 35 | 1.14E-03 | - | 20 | 0.123 | [7] |
| $MoS_2$ | 3.8 | 80 | 8.30E-04 | - | 51 | 0.54 | [8] |
| $MoS_2$ | 4 | 10 | 5.20E-04 | 1.42E-04 | - | - | [9] |
| $MoS_2$ | 5 | 100 | 4.60E-04 | - | 55 | 0.5 | [10] |
| $MoS_2$ | 0.65 | 380 | 7.00E-04 | - | 33.5 | 0.48 | [11] |
| $MoS_2$ | 0.65 | 100 | 3.90E-04 | - | - | 1.1 | [12] |
| $MoS_2$ | 1.95 | 70 | 3.70E-04 | 1.00E-04 | 6 | 1.8 | [13] |
| $MoS_2$ | 0.65 | 500 | 4.50E-04 | - | 102.6 | - | [14] |
| $MoS_2$ | 2.6 | 2000 | 2.71E-04 | - | 22 | 2.2 | [15] |
| $MoS_2$ | 3.3 | 2000 | 2.04E-04 | - | 25 | - | [15] |
| BP | 14.9 | 200 | 1.04E-03 | - | - | - | [16] |
| BP | 12.5 | 100 | 1.20E-03 | - | - | - | [17] |
| BP | 8 | 200 | 8.50E-04 | 3.40E-04 | 144 | 0.58 | [18] |
| $WS_2$ | 1.3 | 100 | 3.10E-04 | 3.20E-04 | 20 | - | [19] |
| $WS_2$ | 2.1 | 40 | 7.00E-04 | - | - | 0.5 | [20] |

# References

1. Li, S. *et al.* Nanometre-thin indium tin oxide for advanced high-performance electronics. *Nat. Mater.* **18**, 1091–1097 (2019).

2. Li, S., Gu, C., Li, X., Huang, R. & Wu, Y. 10-nm Channel Length Indium-Tin-Oxide transistors with $I_{on}$ = 1860 µA/µm, $G_m$ = 1050 µS/µm at $V_{ds}$ = 1 V with BEOL Compatibility. in *IEEE Int. Electron Devices Meet.* 905–908 (IEEE, 2020).

3. Samanta, S. *et al.* Amorphous IGZO TFTs featuring Extremely-Scaled Channel Thickness and 38 nm Channel Length : Achieving Record High $G_{m,max}$ of 125 µS/µm at $V_{DS}$ of 1 V and $I_{ON}$ of 350 µA/µm. in *IEEE Symposium on VLSI Technology* TH2.3 (IEEE, 2020).

4. Han, K. *et al.* First Demonstration of Oxide Semiconductor Nanowire Transistors: a Novel Digital Etch Technique, IGZO Channel, Nanowire Width Down to ~20 nm, and $I_{on}$ Exceeding 1300 µA/µm. in *IEEE Symposium on VLSI Technology* (IEEE, 2021).

5. Matsubayashi, D. *et al.* 20-nm-Node trench-gate-self-aligned crystalline In-Ga-Zn-Oxide FET with high frequency and low off-state current. in *IEEE Int. Electron Devices Meet.* 141–144 (IEEE, 2015).

6. Chakraborty, W. *et al.* BEOL Compatible Dual-Gate Ultra Thin-Body W-Doped Indium-Oxide Transistor with $I_{on}$ = 370µA/µm, SS = 73mV/dec and $I_{on}/I_{off}$ ratio > $4x10^9$. in *Symposium on VLSI Technology* TH2.1 (IEEE, 2020).

7. Shen, P. C. *et al.* Ultralow contact resistance between semimetal and monolayer semiconductors. *Nature* **593**, 211–217 (2021).

8. Liu, Y. *et al.* Pushing the Performance Limit of Sub-100 nm Molybdenum Disulfide Transistors. *Nano Lett.* **16**, 6337–6342 (2016).

9. Yang, L., Lee, R. T. P., Rao, S. S. P., Tsai, W. & Ye, P. D. 10 nm nominal channel length $MoS_2$ FETs with EOT 2.5 nm and 0.52 mA/µm drain current. in *2015 73rd Annual Device Research Conference (DRC)* 237–238 (IEEE, 2015).

10. Lingming Yang *et al.* High-performance $MoS_2$ field-effect transistors enabled by chloride doping: Record low contact resistance (0.5 kΩ·µm) and record high drain current (460 µA/µm). in *IEEE Symposium on VLSI Technology* 192–193 (IEEE, 2014).

11. McClellan, C. J., Yalon, E., Smithe, K. K. H., Suryavanshi, S. V. & Pop, E. High Current Density in Monolayer $MoS_2$ Doped by $AlO_x$. *ACS Nano* **15**, 1587–1596 (2021).

12. Chou, A.-S. *et al.* High On-Current 2D nFET of 390 µA/µm at $V_{DS}$ = 1V using Monolayer CVD $MoS_2$ without Intentional Doping. in *IEEE Symposium on VLSI Technology* TN1.7 (IEEE, 2020).

13. Krasnozhon, D., Dutta, S., Nyffeler, C., Leblebici, Y. & Kis, A. High-frequency, scaled $MoS_2$ transistors. in *IEEE Int. Electron Devices Meet.* 703–706 (IEEE, 2015).

14. Li, T. *et al.* Epitaxial growth of wafer-scale molybdenum disulfide semiconductor single crystals on sapphire. *Nat. Nanotechnol.* (2021).

15. Kang, J., Liu, W. & Banerjee, K. High-performance MoS$_2$ transistors with low-resistance molybdenum contacts. *Appl. Phys. Lett.* **104**, 2–7 (2014).

16. Si, M., Yang, L., Du, Y. & Ye, P. D. Black phosphorus field-effect transistor with record drain current exceeding 1 A/mm. in *2017 75th Annual Device Research Conference (DRC)* (IEEE, 2017).

17. Li, X. *et al.* High-speed black phosphorus field-effect transistors approaching ballistic limit. *Sci. Adv.* **5**, eaau3194 (2019).

18. Yang, L. M. *et al.* Few-layer black phosporous PMOSFETs with BN/Al$_2$O$_3$ bilayer gate dielectric: Achieving I$_{on}$= 850 μA/μm, g$_m$= 340 μS/μm, and R$_c$=0.58 kΩ·μm. in *IEEE Intl. Electron Devices Meet.* 127-130 (IEEE, 2016).

19. Lin, D. *et al.* Scaling synthetic WS$_2$ dual-gate MOS devices towards sub-nm CET. in *IEEE Symposium on VLSI Technology* (IEEE, 2021).

20. Pang, C.-S., Wu, P., Appenzeller, J. & Chen, Z. Sub-1nm EOT WS$_2$ -FET with I$_{DS}$ > 600μA/μm at V$_{DS}$ =1V and SS < 70mV/dec at L$_G$ =40nm. in *IEEE Int. Electron Devices Meet.* 43–46 (IEEE, 2020).